\documentstyle[preprint,aps]{revtex}

\newcommand{\doublespace}{
  \renewcommand{\baselinestretch}{1.6}\large\normalsize}

\begin{document}

\hyphenation{mono-pole mono-poles lat-tice}
\newcommand{\w}{\left<W(R,T)\right>}
\newcommand{\wmon}{W_{mon}^e}
\newcommand{\wph}{W_{ph}^e}
\newcommand{\ts}{\textstyle}

\begin{titlepage}

\begin{tabbing}
\` ILL-(TH)-94-\#14 \\
\` April, l994 \\
\end{tabbing}

%\vspace*{.7in}

\begin{center}
{\bf String Tension from Monopoles \\
in $SU(2)$ Lattice Gauge Theory \\}
%latex file: su2mono.tex\\}
\vspace*{.5in}
John D. Stack and Steven D. Neiman\\
\vspace*{.2in}
{\it Department of Physics \\
University of Illinois at Urbana-Champaign \\
1110 W. Green Street \\
Urbana, IL 61801 \\}
\vspace*{.2in}
and \\
\vspace*{.2in}
Roy J. Wensley \\
\vspace{.2in}
{\it Department of Mathematical Sciences \\
Saint Mary's College \\
Moraga, CA 94575 \\
}
\vspace*{0.3in}
{\it (submitted to Physical Review D)}
\end{center}

\begin{tabbing}
\` PACS Indices: 11.15.Ha\\
\` 12.38.Gc\\
\end{tabbing}

%\hspace{2.0in}
%\begin{flushright}
%\today
%\end{flushright}
\end{titlepage}
\vfill\eject

\doublespace
\pagestyle{empty}

\begin{center}
{\bf Abstract}
\end{center}

\noindent We calculate the heavy quark potential from the magnetic current
due to monopoles in
four dimensional $SU(2)$ lattice gauge theory.  The magnetic current is
located in
configurations generated in a conventional Wilson action simulation
on a $16^4$ lattice.  The configurations
are projected with high accuracy into the maximum abelian gauge.  The
magnetic current is then extracted and the
monopole contribution to the potential is calculated.
The resulting string tension is in excellent agreement with the $SU(2)$ string
tension obtained by conventional means from the configurations.  Comparison
is made with the $U(1)$ case, with emphasis on the differing
periodicity properties of $SU(2)$ and $U(1)$ lattice gauge theories.
The properties of the maximum abelian gauge are discussed.

\vspace*{.5in}

\newpage
\pagestyle{plain}

\section{Introduction }
\label{sec-intro}
	In this paper, we report on our calculations of the $SU(2)$ string
tension using monopoles.  The monopoles were located in $SU(2)$ lattice
gauge theory configurations after the configurations were projected with high
accuracy into the maximum abelian gauge.
These lattice calculations were motivated by an approach to continuum
confinement outlined by 't Hooft some time ago~\cite{thoo}.  In 't Hooft's
framework,
the first step is a partial gauge-fixing, applied only to those gauge fields
which are ``charged", or have off-diagonal generators in the Lie algebra of
the gauge group.
 The central idea is that the
monopoles associated with the abelian gauge invariance left after
partial gauge-fixing will control
non-perturbative phenomena.
This abelian
gauge invariance is associated
with the fields whose generators are diagonal,   called ``photons".
For an $SU(2)$ gauge group with the generator $T_3$ diagonal, the
gauge field $A^{3}_{\mu}$ is the abelian field or photon, and gauge-fixing
is done only on $A^{a}_{\mu}, a=1,2$, or equivalently the charged fields
\begin{displaymath}
W^{\pm}_{\mu} = \frac{1}{\sqrt{2}}(A^{1}_{\mu} \pm iA^{2}_{\mu}) .
\end{displaymath}

For the particular choice of gauge-fixing condition known as the maximum
abelian gauge, the continuum functional
\begin{equation}
G_{c}\equiv \frac{1}{2}\sum_{\mu}\int \left ( (A^{1}_{\mu})^{2}
+(A^{2}_{\mu})^{2} \right )d^{4}x
=\sum_{\mu}\int(W^{+}_{\mu}W^{-}_{\mu})d^{4}x
\label{contgf}
\end{equation}
is minimized over all $SU(2)$ gauge transformations, leading to the
conditions
\begin{displaymath}
(\partial_{\mu}+igA^{3}_{\mu})W^{+}_{\mu}=
(\partial_{\mu}-igA^{3}_{\mu})W^{-}_{\mu}=0,
\end{displaymath}
where $g$ is the $SU(2)$ gauge coupling.
  In the remainder of this Section, we discuss how
this is turned into a specific calculational scheme on the lattice.
Our results are described in Section~\ref{sec-mono} .  Section~\ref{concl}
contains a discussion of the maximum abelian gauge and our conclusions.

\subsection{~Lattice Gauge-Fixing}~The $SU(2)$ lattice gauge theory is built
out of link variables
$U_{\mu}(x)$,
\begin{displaymath}
U_{\mu}(x)\equiv e^{iga\vec{A}_{\mu}\cdot\vec{\tau}},
\end{displaymath}
where $\vec{\tau}=\vec{\sigma}/2$ are the generators of $SU(2)$ in the
fundamental representation, and $a$ is the lattice spacing.  On the lattice,
the maximum abelian gauge is obtained by maximizing the lattice functional
\begin{equation}
G_{l}=\sum_{x,\mu}\frac{{\rm
tr}}{2}\left[U^{\dagger}_{\mu}(x)\sigma_{3}U_{\mu}(x)
\sigma_{3}\right]
\label{lattgf}
\end{equation}
over all $SU(2)$ gauge transformations~\cite{kron}.  It is easy to show that
in the continuum limit,
maximizing $G_{l}$ is equivalent to
minimizing $G_{c}$.

At the maximum, $G_{l}$ will be stationary under gauge transformations.
The demand that $G_{l}$ be stationary with respect to a gauge transformation
at  an arbitrary site $y$ leads to the requirement that
\begin{equation}
X(y)\equiv \sum_{\mu}\left[U_{\mu}(y)\sigma_{3}U^{\dagger}_{\mu}(y)
+U^{\dagger}_{\mu}(y-\hat{\mu})\sigma_{3}U(y-\hat{\mu})\right]
\label{Xdef}
\end{equation}
be diagonal.  This can be accomplished by a gauge transformation
$\Omega(y)$.  However, the value of $X$ at the nearest neighbors of $y$ is
affected by $\Omega(y)$, so the diagonalization of $X$ over the whole lattice
must be done iteratively.

After gauge-fixing, there is still manifest $U(1)$ gauge invariance, and
it is useful to factor a $U(1)$ link variable from $U_{\mu}(x)$, writing
$
U_{\mu}(x)=u_{\mu}(x)w_{\mu}(x),
$
where
$u_{\mu}(x)=\exp(i\phi_{\mu}^{3}\tau_{3})$, and
$w_{\mu}(x)=\exp(i\vec{\theta_{\mu}}\cdot\vec{\tau}),$
with $\theta^{3}_{\mu}\equiv0$.
The $U(1)$ gauge transformation properties of $u_{\mu}$ and $w_{\mu}$
follow upon applying an abelian
transformation $\Omega_{3}=\exp(i\alpha(x)\tau_{3})$ to
$U_{\mu}$:
$$
u_{\mu}(x)\rightarrow \Omega^{\dagger}_{3}(x+\hat{\mu})u_{\mu}(x)\Omega_{3}(x),
$$

$$
w_{\mu}(x)\rightarrow \Omega^{\dagger}_{3}(x)w_{\mu}(x)\Omega_{3}(x),
$$
so $w_{\mu}$ transforms as a charged chiral field at $x$.
The angle $\phi^{3}_{\mu}$ can be extracted
from the matrix elements of $U_{\mu}$
by expanding the gauge-fixed $SU(2)$
link $U_{\mu}$ in Pauli matrices, writing
\begin{equation}
U_{\mu}=U_{\mu}^{0}+i\sum_{k=1}^{3}U_{\mu}^{k}\cdot \sigma_{k}.
\end{equation}
Then $\phi^{3}_{\mu}=2\cdot\arctan(U_{\mu}^{3}/U_{\mu}^{0})$.

	The maximum abelian gauge globally suppresses $|\vec{\theta}_{\mu}|$,
or equivalently, tries to force $w_{\mu}$ to the identity matrix.  Even
so, it is non-trivial to expect that long range effects associated with
confinement are totally isolated in $u_{\mu}$.
The first concrete calculations to
test this were performed by Suzuki and Yotsuyanagi~\cite{suz1}.
 In Wilson loops, they
replaced each $SU(2)$ link variable by the $U(1)$ link variable $u_{\mu}$,
and found
that full $SU(2)$ results were obtained for Creutz ratios.
This did not work for forms of partial gauge-fixing other than
the maximum abelian gauge.
\subsection{~Monopoles in U(1) }
In $U(1)$ lattice gauge theory, the usual form of a Wilson loop
involves a line integral of the $U(1)$ link variable $\phi_{\mu}(x)$ taken
around a path specified by an integer-valued current $J_{\mu}(x)$:
\begin{equation}
\left< W_{U(1)} \right>=
\left<
\exp \left(i\sum_{x}\phi^{3}_{\mu}J_{\mu}\right)\right>,
\label{u1basic}
\end{equation}
where $\left< \cdot \right>$ denotes the expectation value over
the ensemble of $U(1)$ configurations.
In addition, it is well established that  confinement  in $U(1)$ is via
monopoles~\cite{js7,js7-2}.
A  Wilson
loop originally expressed as in Eq.(\ref{u1basic})
can be factored into a
perturbative term arising from one photon exchange, times a
non-perturbative term arising from monopoles,\footnote{
Eq.(\ref{u1fac}) can be derived as an exact formula
only for the Villain form of the $U(1)$ action.  However, in Ref.~\cite{js7},
it was shown to work for other forms of the action, provided a coupling
constant
mapping was used to calculate $\left<W_{phot}\right>$.}
\begin{equation}
\left< W_{U(1)} \right> = \left< W_{phot} \right> \cdot
\left< W_{mon} \right>.
\label{u1fac}
\end{equation}

An explicit formula for $\left< W_{mon} \right>$  in $U(1)$
is obtained by writing
$J_{\mu}$ as the curl of
a Dirac sheet variable~\cite{diracsh};
$J_{\mu}=\partial_{\nu}D_{\mu\nu}$,
where $\partial_{\nu}$ denotes a discrete derivative.
Then $\left< W_{mon} \right>$ is given by
\begin{equation}
\left< W_{mon} \right>=
                \left<\exp\left(\frac{i2\pi}{2}\sum_{x}
D_{\mu\nu}(x)
                   F^{*}_{\mu\nu}(x)\right)\right>_m ,
\label{eqnmon}
\end{equation}
where
$\left<\cdot\right>_m$ denotes the sum over configurations of magnetic current.
The sheet variable $D_{\mu\nu}$
is not unique. For  the usual case of an $R\times T$ loop with $|J_{\mu}|=1$,
a useful choice is to set
$D_{\mu\nu}=1$ on the
plaquettes of the flat rectangle
with boundary  $J_{\mu}$, and $D_{\mu\nu}=0$ on all other
plaquettes.
In Eq.(\ref{eqnmon}), $F^{*}_{\mu\nu}$ is the dual of the field strength
due to the magnetic current;
%% FOLLOWING LINE CANNOT BE BROKEN BEFORE 80 CHAR
\mbox{$F^{*}_{\mu\nu}(x)=\frac{1}{2}\epsilon_{\mu\nu\alpha\beta}F_{\alpha\beta}(x)$.}
The field strength itself is derived from a magnetic vector potential
$A^{m}_{\mu}$,
\mbox{$F_{\mu\nu}\equiv\partial_{\mu}A^{m}_{\nu}-\partial_{\nu}A^{m}_{\mu}$,}
where
\begin{equation} A^{m}_\mu(x)=
\sum_{y}v(x-y)m_{\mu}(y),
\label{eqnamag}
\end{equation}
and $m_{\mu}$ is the integer-valued, conserved magnetic current.

In Ref.~\cite{suz1}, results characteristic of confinement in $SU(2)$
were obtained
by replacing the full Wilson loop by a $U(1)$ loop expressed in terms
of the link variable $\phi^{3}_{\mu}$.  Since confinement in $U(1)$ lattice
gauge theory itself is via monopoles, this raises the possibility of
a monopole explanation of confinement in $SU(2)$.

\subsection{~Periodicity in SU(2)}~Before applying Eqs.(\ref{eqnmon})
and
(\ref{eqnamag}) to $SU(2)$, account must be taken
of the difference in periodicity for
the standard $U(1)$ and $SU(2)$ actions.
{}~The above discussion was for a
$U(1)$ action periodic in link angles
with period $2\pi$.  The $SU(2)$ action is periodic in the $U(1)$ link
angle $\phi_{\mu}^{3}$ with period $4\pi$. This follows from the formula
$u_{\mu}(x)=\exp(i\phi_{\mu}^{3}\tau_{3})$, where $\tau_3=\sigma_3/2$.
Alternatively,
the process of temporarily setting $w_{\mu}$
to the identity
on every link transforms the $SU(2)$ action into a $U(1)$ action with period
$4\pi$:
\begin{equation}
S \longrightarrow
\frac{\beta}{2}\sum_{x,\mu>\nu}\left(1-\cos(\frac{\phi^{3}_{\mu\nu}(x)}{2})
\right),
\label{u1approx}
\end{equation}
where $\phi^{3}_{\mu\nu}=\
\partial_{\mu}\phi^{3}_{\nu}-\partial_{\nu}\phi^{3}_{\mu}$.

While
Eq.(\ref{u1approx}) is not intended as an approximation to the
full $SU(2)$ action, it does correctly reveal the monopole charges
which will occur in
$SU(2)$.  A Dirac string occurs when the plaquette angle $\phi^{3}_{\mu\nu}$ is
an integer multiple of $4\pi$, rather than $2\pi$, so the magnetic current
$m_{\mu}$ here is
an even integer, or in other words the abelian  monopoles which occur in
$SU(2)$ are Schwinger monopoles~\cite{schwinger,thoo}.

The replacement of $w_{\mu}$ by the identity in the action
as in Eq.(\ref{u1approx}) is never actually done.
The configurations are generated using full $SU(2)$ dynamics.
However, after projecting the configurations into the
maximum abelian gauge, the $w_{\mu}$ are set to the identity on every
link in the calculation of Wilson loops. Only the $U(1)$ link variable
$u_{\mu}(x)$ is retained,  so that the $SU(2)$ Wilson loop becomes
a $U(1)$ loop:

\begin{equation}
W_{SU(2)}\longrightarrow \frac{1}{2} \times
\left< \left[
\exp \left(i\sum_{x}\phi^{3}_{\mu}J_{\mu}\right)
+\exp \left(-i\sum_{x}\phi^{3}_{\mu}J_{\mu}\right)
\right] \right>,
\label{wu1approx}
\end{equation}
where the conserved line current $J_{\mu}$ has $|J_{\mu}|=\frac{1}{2}$.
Since in this approximation, the
Wilson loop is built out of $U(1)$ variables which involve
only $\phi^{3}_{\mu}$,
perturbative exchange of gluons coupled
to $\tau_{1}$ and $\tau_{2}$ has clearly been suppressed.
By analogy with the situation in $U(1)$,
perturbative exchange of the neutral gluon
or ``photon"
coupled to $\tau_{3}$ is still allowed, but is expected to reside in the
$SU(2)$ analog of the
$\left<W_{phot}\right>$ factor of Eq.(\ref{u1fac}), whereas
the confining part of the potential
is expected to reside in the factor $\left<W_{mon}\right>$.  That the confining
potential
resides solely in $\left<W_{mon}\right>$ is a postulate
which will be justified by our results.

Since $|J_{\mu}|=\frac{1}{2}$,
the calculation of $\left<W_{mon}\right>$
for $SU(2)$
involves a Dirac sheet variable with $|D_{\mu\nu}|=\frac{1}{2}$,
along with a magnetic
current
$m_{\mu}$ which is an even integer.  Both of these arose from the $4\pi$
periodicity of $SU(2)$ in the link angle $\phi^{3}_{\mu}$. It is
straightforward
to transform back to the familiar case of $2\pi$ periodicity.
Define $\bar{\phi}^{3}_{\mu}=\phi^{3}_{\mu}/2
=\arctan(U^{3}_{\mu}/U^{0}_{\mu})$, and require $\bar{\phi}^{3}_{\mu}
\in (-\pi,\pi]$.
The location of the magnetic current  starts with plaquette
angles $\bar{\phi}^{3}_{\mu\nu}$ constructed from
$\bar{\phi}^{3}_{\mu}$,
\begin{equation}
\bar{\phi}^{3}_{\mu\nu}(x)=
\partial_{\mu}\bar{\phi}^{3}_{\nu}
-\partial_{\nu}\bar{\phi}^{3}_{\mu}=
\bar{\phi}^{3}_{\mu}(x)
+\bar{\phi}^{3}_{\nu}(x+\hat{\mu})-\bar{\phi}^{3}_{\mu}(x+\hat{\nu})
-\bar{\phi}^{3}_{\nu}(x)
\end{equation}
The plaquette angle $\bar{\phi}^{3}_{\mu\nu}$ is resolved into a
Dirac string contribution, plus a fluctuating part:
\begin{equation}
\bar{\phi}^{3}_{\mu\nu}=2\pi \bar{n}_{\mu\nu}+\tilde{\phi}^{3}_{\mu\nu},
\label{eqnphimunu}
\end{equation}
where
$\tilde{\phi}^{3}_{\mu\nu}\in (-\pi,\pi]$
and $\bar{n}_{\mu\nu}$ is an integer~\cite{degrand}.
The integer-valued magnetic current $\bar{m}_{\mu}=m_{\mu}/2$ is determined by
the net flux
of Dirac strings into an elementary cube:
\begin{equation}
\bar{m}_{\mu}
=-\frac{1}{2}\epsilon_{\mu\nu\alpha\beta}
\cdot\partial_{\nu} \bar{n}_{\alpha\beta}.
\label{eqnnmunu}
\end{equation}
We now simply use $\bar{m}_{\mu}$ in Eq.(\ref{eqnamag})
to obtain $A_{\mu}$, and from it,
$F^{*}_{\mu\nu}$.  Finally $\left<W_{mon}\right>$ for $SU(2)$ is obtained from
Eq.(\ref{eqnmon}) where the Dirac sheet variable now has $|D_{\mu\nu}|=1$.
We have in effect used
the Dirac condition~\cite{dirac} on
the product of electric
and magnetic charge to transform from $4\pi$ to $2\pi$
periodicity .
In the interaction of a
distribution of magnetic current with a single electric charge, if the
magnetic current is halved and the electric current is doubled, the same result
is obtained,
since the interaction only depends on the quantized product of electric
and magnetic charge.

\subsection{~Monopole Location in 1-Cubes}~To summarize, we use
\label{sec-monoloc}
Eq.(\ref{eqnmon}) to calculate $\left<W_{mon}\right>$ for
$SU(2)$, after
extracting the magnetic current $\bar{m}_{\mu}$ from $SU(2)$ configurations
projected into the maximum abelian gauge.  In the
identification of $\bar{m}_{\mu}$, elementary cubes were used, so
a magnetic charge is located at the center of a spacial 1-cube,
likewise for other components of the magnetic current.  This is done
mainly for practical reasons; if (say) 2-cubes
were used instead, the effective lattice size would become $8^4$ instead of
$16^4$.  Although this practical reason was dominant in our
calculations,  it is worth pointing out that
there is no conflict between
monopoles being extended objects, and using 1-cubes to locate them.
The monopole location procedure finds a monopole by
locating the end of a Dirac string.  Therefore, we must be in  a gauge where
there are Dirac strings attached to monopoles.
Consider for the moment the
case of a 't Hooft-Polyakov monopole in the continuum~\cite{thoopolymon}.
The gauge where a Dirac string  is present is usually described
as having the
Higgs field along the isotopic 3-axis.  A completely equivalent
description is that the gauge fields
are in the maximum abelian gauge .
In this gauge, the photon field $A^{3}_{\mu}$ takes the form appropriate
for an elementary point monopole.
The
extended structure of the monopole involves only the charged gauge fields,
$W_{\mu}^{\pm}$~\cite{smit}.
In a gauge with these properties,
the center of the monopole may be located by finding the end of
its Dirac string, using the finest spacial scale available.

In calculating
the {\em effects} of monopoles on heavy quarks, the assumption is made that
despite their finite size,  only the long range Coulombic fields
produced by monopoles contribute to the confining potential.
This produces a clear calculational procedure for computing
$\left<W_{mon}\right>$,
which as will be seen in the next Section, works very well. However, further
theoretical understanding of this assumption is certainly needed.

\section{Monopole Calculations}
\label{sec-mono}
\subsection{Simulation and Gauge-Fixing}

	Our simulations were done on a $16^4$ lattice, using the standard
Wilson form of the $SU(2)$ action.  Three $\beta$ values were used;
$\beta=2.40,2.45$, and $2.50$.  At each $\beta$, after equilibration, $500$
configurations were saved.  Saved configurations were separated by $20$ updates
of the lattice, where a
lattice update consisted of one heatbath sweep~\cite{creutz},
plus one or two overrelaxation sweeps~\cite{overr}.
%need overrelaxation reference
Each of these configurations was
then projected into the maximum abelian gauge using the overrelaxation
method of
Mandula and Ogilvie, with their parameter $\omega=1.70$~\cite{mandula}.
The overrelaxation process was stopped after the off-diagonal elements
of $X(x)$ of Eq.(\ref{Xdef}) were sufficiently small.  Expanding $X(x)$
in Pauli matrices,
\begin{equation}
X=X^{0}(x)+i\sum_{k=1}^{3}X^{k}(x)\cdot \sigma_{k},
\end{equation}
we used
\begin{equation}
\left<|X^{ch}|^{2}\right>\equiv \frac{1}{ L^{4}}\sum_{x}\left(
|X^{1}(x)|^{2}+|X^{2}(x)|^{2}\right)
\end{equation}
as a measure of the average size of the off-diagonal matrix elements of
X over the lattice,
and required $\left<|X^{ch}|^{2}\right>\le 10^{-10}$.  This condition was
reached
in approximately 1000 overrelaxation sweeps.
%Remark on how severe is this condition.

{}From each gauge-fixed $SU(2)$ link, the $U(1)$
link angle $\bar{\phi^{3}_{\mu}}$ was extracted using the formula
$\bar{\phi}^{3}_{\mu} =\arctan(U^{3}_{\mu}/U^{0}_{\mu})$,
as described in   Section~\ref{sec-intro}.  Then making use of the plaquette
angles
$\bar{\phi}^{3}_{\mu\nu}$,
the magnetic current
$\bar{m}_{\mu}(x)$ was found.  Although the values
$\bar{m}_{\mu}=0\pm1\pm2$ are allowed by Eqs.(\ref{eqnphimunu}) and
(\ref{eqnnmunu}), the overwhelming fraction of links carrying current
had $|\bar{m}_{\mu}|=1$; $|\bar{m}_{\mu}|=2$ rarely occured.
Only a few percent of the links actually carried current.
We define $\rm{f}_{m}$ as the
number of links with non-zero current, divided by the total number of links,
$4\cdot L^{4}$.  Our results for $\rm{f}_{m}$ are recorded in Table I, and
agree well with those in the literature~\cite{borny,debbio}.

\subsection{The Quark Potential from Monopoles}

With the magnetic current $\bar{m}_{\mu}$ in hand, the monopole Wilson
loops for $SU(2)$ are completely determined. Applying
Eqs.(\ref{eqnmon}) and (\ref{eqnamag}), for $\beta=2.40$ and $2.50$
all $R\times T$ loops up to $7 \times 10$
were measured and averaged over configurations. For $\beta=2.45$,
the maximum size was increased to $8 \times 12$.
{}From the monopole Wilson loops, monopole contributions to the
potential, denoted by $V_{mon}(R)$,
were extracted  by performing straight line
fits to $\ln(\left<W(R,T)_{mon}\right>)$ vs $T$.  The fits for $R \ge 2$
were over the interval $T=R+1$ to
$T_{max}$, except for $R_{max}$, where $T=R$ to $T_{max}$ was used.
Finally, the monopole contribution to the string tension $\sigma$ was extracted
by fitting $V_{mon}(R)$ to the form  $V_{mon}(R)=\alpha/R+\sigma\cdot R+V_{0}$,
over
the interval $R=2$ to $R=R_{max}$.  The results are shown in
Fig.~(1), and tabulated
in Table II.

As a glance at either Table II or Fig.~(1) shows, the monopole potentials
are essentially linear at all values of R, with  negligible Coulomb terms.
This is expected on the basis of the discussion of Section~\ref{sec-intro}.
The steps of first
setting $w_{\mu}=1$ on each link of the Wilson loop, and then
calculating only the monopole factor $\left<W_{mon}\right>$ of the resulting
$U(1)$
loop, effectively suppresses the Coulomb terms arising from single gluon
exchange.

The crucial question for the present paper is whether or not the
monopole string tensions agree with those for full $SU(2)$.
Of the various ways of determining the $SU(2)$ string tension, the
``torelon" method of Michael and Teper is perhaps the best
comparison.  The torelon method involves the temporal correlation
between loops which are wrapped around
the entire lattice in the spacial direction. The method determines the string
tension directly, without the need to separate linear from Coulomb terms.
For $\sqrt{\sigma}$, Michael and Teper give 0.258(2) at $\beta=2.4$ on a
$16^4$ lattice and 0.185(2) at $\beta=2.5$
on a $20^4$ lattice~\cite{michael}.  Both
numbers are in excellent agreement with the monopole results of Table II.
As an additional check that full $SU(2)$ results were being obtained,
we performed linear-plus-Coulomb fits to
our own $SU(2)$ Wilson loops,
which were obtained from the same set of configurations using an analytic
form of the multi-hit method~\cite{parisi}, along with the smearing
method~\cite{smear}, as noise reduction techniques.
The resulting string tension
is presented in Table III, and again there is agreement with the monopole
string tension to within statistical
errors.  Statistical errors were estimated using standard jackknife methods.
Similar results for the $SU(2)$ string tension using monopoles
have recently been obtained by
Shiba and Suzuki~\cite{suz3}. The Coulomb terms coming from the monopoles
and full $SU(2)$ naturally differ, as explained above.  No attempt to calculate
the $SU(2)$ analog of the factor $\left<W_{phot}\right>$ is made in this paper.

\subsection{The String Tension and Magnetic Current Loop Size}

To further investigate confinement via monopoles, the magnetic current was
resolved into individual loops, each of which separately conserves current.
Not all sizes of loops are expected to contribute to the confining
potential.  For example in $U(1)$ lattice gauge theory,
it is possible to show analytically that a random distribution of
loops much smaller in
size than the Wilson loop under consideration does not affect the long range
potential, but only renormalizes the Coulomb term.
At the opposite extreme, again
in $U(1)$, it is known that the string tension can be
calculated accurately using only the contribution coming from
very large loops of magnetic current~\cite{js-rw loops}.

For the present case of $SU(2)$, we first obtained a rough measure of
how the current is distributed over loops of various size.  This was done
by counting the number of current-carrying links residing in loops of size
up to and including
$ 10,20,50, {\rm and} ~100$  links.
In Table~IV,  the results are shown as a fraction of the total current.
As can be seen from the last column of the table,
including all loops of magnetic current with up to 100 links
accounts for approximately half of the current-carrying links at
each coupling.  The remainder of the current consists of a small number
($\sim 3$) of loops, each containing typically several hundred links.

To study how the string tension depends on magnetic current
loop size,
we computed $\left<W_{mon}\right>$ with a cut on loop size.  Since experience
with $U(1)$
shows that large loops are what is important in confinement, in
computing $\left<W_{mon}\right>$ from
Eq.(\ref{eqnmon}),
only those loops of magnetic current with more than $ n_{cut}$ links were
included.
The results for the string tension with $n_{cut}=50 ,100, {\rm and} ~200$
are shown in Table~V.  For $n_{cut}=50$ the answers are within
statistical errors of the full string tension for all couplings.  However,
as $n_{cut}$ is increased the string tension decreases steadily so that by
$n_{cut}=200$, there is a statistically significant deviation from the full
string tension.
This shows that
in $SU(2)$ lattice gauge theory, the string tension cannot be explained by
retaining only the
very largest loops of current, unlike the situation in $U(1)$ .
To make this point quantitatively, we compare
to our work
in $U(1)$ lattice gauge theory on a $24^4$ lattice.
In Ref.~\cite{js-rw loops}, we considered a $U(1)$ coupling corresponding to
a string tension of
$\sigma=0.058(2)$, intermediate between the $SU(2)$ string tension
at $\beta=2.40$ and $\beta=2.45$.  For this case and other nearby couplings,
the $U(1)$ string tension is stable under an increase of $n_{cut}$ to at least
$n_{cut}=1,000$, whereas deviations are already significant
in $SU(2)$ for $n_{cut}=200$.  Admittedly, two different lattice sizes are
being
compared here, but since our $16^4$   $SU(2)$ string tension
is within statistical errors of
$24^4$  $SU(2)$ numbers,
it is reasonable to assume that the distribution of
loops we found on $16^4$ would be similar to that on a
$24^4$ lattice.

 Small loops of magnetic
current play no role in confinement for either $SU(2)$ or $U(1)$.
In the present work, we performed additional calculations of
$\left<W_{mon}\right>$ with a cut on loop size, but this time including only
loops of
current with less than $ n_{cut}$ links.
The string tension
was statistically zero for $n_{cut}=50$ and 100 for all three values
of $\beta$.  The
Coulomb terms were small,
of magnitude $\approx$ 10\% of the Coulomb term for full $SU(2)$, and
attractive in sign.
For $n_{cut}=200$, a very small contribution to the
string tension was seen for $\beta=2.5$.

We may summarize our investigation of current loops
by saying that for
$SU(2)$, somewhat more
than
half the magnetic
current plays a role in the confining potential.
The string tension
can be explained by retaining all magnetic current loops of intermediate size
($\approx 50$ links)
and larger.
An reasonable guess for how large a loop must be
to play a role in the confining potential would be a physical extent
 $O(1/\sqrt{\sigma})$ or greater.  For $U(1)$, there is a class of
huge magnetic current loops present only in the confined phase, and the
string tension can be explained with
a restriction only to these very large
loops.

\subsection{Abelian Wilson Loops}

	Another possible route to the string tension is Eq.(\ref{wu1approx}),
which expresses the assumption that after partial gauge-fixing, the physics
of confinement is contained in the $U(1)$ link variable $\phi^{3}_{\mu}$.
   So far, we have concentrated on  the extraction of the string tension from
$\left<W_{mon}\right>$, which is based on the further assumption that the
$U(1)$ approximation to the full $SU(2)$ Wilson loop, Eq.(\ref{wu1approx}), can
be factored into
a photon part $\left<W_{phot}\right>$, times a  monopole part
$\left<W_{mon}\right>$.
The physical motivation for this was to investigate the monopole
confinement mechanism,
but the method also has clear computational
advantages.  The noise level in the monopole Wilson loops is much lower
than in the full $SU(2)$ loops, and the string tension is easier to
identify, due to the very small Coulomb terms in the monopole contribution
to the potential.  For completeness, we have returned to Eq.(\ref{wu1approx})
to calculate the $U(1)$ approximation to the $SU(2)$ Wilson loops directly
in terms of the link variable $\phi^{3}_{\mu}$, and
extracted the string tension and Coulomb term.   The statistical accuracy of
these results is poor compared to those obtained from the full $SU(2)$ Wilson
loops, due to the absence of the multi-hit technique.  Partial gauge-fixing
produces a sequence of $U(1)$ configurations, but the $U(1)$ action that would
produce these configurations is unknown.  Lacking the usual
noise reduction techniques, we perform a simple
average over
these $U(1)$ configurations to obtain Wilson loops, and extract the string
tension and Coulomb term from the resulting potential.  The Coulomb term is now
significant, but still less than that for full $SU(2)$.  This is expected
since exchange of neutral gluons or photons is allowed in Eq.(\ref{wu1approx}).
The string tension is consistent with the full $SU(2)$ string tension as well
as that extracted from $\left<W_{mon}\right>$~\cite{suz2}.
In Fig.~(2), we show the potentials
derived from the full $SU(2)$ loops, from the monopole loops, and
from Eq.(\ref{wu1approx}), for $\beta=2.45$.  A constant has been added
to the monopole and $U(1)$ potentials for the purpose of comparing
the $R$ dependence of the three potentials in Fig.~(2).  The shift of
the potentials by a constant is
not physically relevant
%\newpage \noindent
in that it does not affect
the functional
dependence on $R$ of the potentials.  The plots for $\beta=2.40$ and
2.50 are similar.
% Need statement about constant shift either here or in figure caption.

\section{Conclusions and Summary}
\label{concl}

	The present work was done entirely in the maximum abelian gauge,
stimulated by the pioneering work of Suzuki and Yotsuyanagi~\cite{suz1}.
In an average way over the lattice, the maximum abelian gauge forces the
fluctuations in the charged sector to be as small  as possible.
This gauge also has a variational formulation, (Eqs.(\ref{contgf})
and (\ref{lattgf})), and
is renormalizable in the continuum limit~\cite{min}.  While these are
desirable features, it should be possible to
capture the physics of confinement with other forms of partial gauge-fixing.
That this has not been possible so far in lattice calculations may
have to do with
the monopole location procedure.
As mentioned in Section~\ref{sec-monoloc}, the location procedure
finds a monopole by locating the end of its Dirac string, so if monopoles
are to be correctly located using the abelian flux over
1-cubes, the gauge which is used must
attach a string to a monopole at distance scales right down to the lattice
spacing.  For a 't Hooft-Polyakov
monopole in the continuum,
the requirement that the string go all the way to the center of the monopole
singles out the maximum abelian gauge.  It is easy to check
that another gauge discussed by 't Hooft~\cite{thoo},
where the charged field strength $G^{\pm}_{12}$ is set to
zero, does not have this property.  Starting from the ``hedgehog" form of
the monopole solution, going to the gauge where $G^{\pm}_{12}=0$
will be accomplished by  the same gauge transformation as going to
the maximum abelian gauge
in the region far from the monopole, so here there is no distinction between
the
two gauges.
However, inside the extended structure of the monopole,
the two begin to deviate and only the maximum abelian gauge has a Dirac
string extending to
the origin of the monopole.\footnote{At the center of the monopole, the
hedgehog solution already satisfies $G^{\pm}_{12}=0$.}
Assuming a similar phenomenon occurs on the lattice,
a calculation
which uses 1-cubes to locate the monopoles in the lattice version of the
 gauge $G^{\pm}_{12}=0$
may then locate the monopoles incorrectly.
This interpretation of the difference
between the maximum abelian gauge and other forms of partial gauge-fixing
is supported by work on three-dimensional $SU(2)$ lattice gauge theory by
Trottier, {\it et al.}~\cite{trottier}.  They show that the density of
monopoles
as found in the maximum abelian gauge has the correct scaling law as a function
of coupling, using 1-cubes to locate the monopoles.  However, for other forms
of partial gauge-fixing such as $G^{\pm}_{12}=0$, the correct scaling law
is obtained only after the
size of the cube used to locate the monopoles is increased considerably.  This
would say that while other forms of partial gauge-fixing are
in principle equivalent,
for practical
reasons having to do with lattice size
the maximum abelian gauge is likely to continue to be favored
in
calculations with monopoles.

	We have presented evidence in favor of 't Hooft's picture of
confinement.  Although our lattice size and number of configurations
is moderate, the numbers obtained from monopoles for the string tension
are in excellent agreement with full $SU(2)$ numbers on larger lattices
with higher statistics.  It would be of interest to repeat the present
work on a larger lattice with several thousand configurations, in order
to search for any possible systematic difference between the full $SU(2)$
string tension and the string tension deduced from monopoles.  A larger
lattice would also allow a move toward weaker coupling and a smaller string
tension.  As the correlation length defined by the string tension grows,
so presumably must the physical size of monopoles which has been ignored in
the present work.  An interesting issue is whether the string tension will
continue to be dominated by the long range Coulombic fields of the monopoles.
Study of the gluon propagator offers a different line of attack on this
question.
We are presently calculating both charged and neutral gluon
propagators  from our $16^4$ configurations, after a final gauge-fixing which
puts the photon link variable
$\phi^{3}_{\mu}$
in the lattice Landau gauge~\cite{mandulaph}.  A detailed report will be
presented
elsewhere, but our preliminary calculations show that the neutral gluon or
photon propagator is intrinsically larger and of longer range than the
charged gluon propagator.

This work was supported in part by the National Science Foundation under
Grant No. NSF PHY 92-12547.  The calculations were carried out on the Cray
Y-MP system at the National Center for Supercomputing Applications at the
University of Illinois, supported in part by the National Science
Foundation
under Grant No. NSF PHY920026N. R.\ J.\ W.\ would like to acknowledge
support from the Faculty Development Funds of Saint Mary's College of
California.

%bibwr.tex

%
%   BIBLIOGRAPHY FILE
%

\newpage

\section*{Captions}

\noindent {\bf Table I}.  The fraction of links carrying magnetic current
at each value of $\beta$.

\vspace{.15in}

\noindent {\bf Table II}.  The string tension and Coulomb coefficient
obtained from fits to the potentials
calculated using the monopole Wilson loops at each $\beta$.

\vspace{.15in}

\noindent {\bf Table III}.  The  string tension and Coulomb coefficient
for the full SU(2) potential at each $\beta$.

\vspace{.15in}

\noindent {\bf Table IV}. The fraction of total magnetic current
contained in loops with
$\le$ 10,20,50, and 100 links.

\vspace{.15in}

\noindent {\bf Table V} The monopole string tension from all current loops with
$\ge$ 50,100, and 200 links.

\noindent {\bf Figure 1}.  The potentials extracted from monopole Wilson
loops at $\beta=2.40$ (circles), 2.45 (squares), and 2.50 (triangles).
The solid lines are the linear-plus-Coulomb fits to each potential.

\vspace{.15in}

\noindent {\bf Figure 2}.  Comparison of the monopole potentials (circles),
the $U(1)$ approximation to full $SU(2)$ potentials (squares) and the
full $SU(2)$ potentials (triangles) at $\beta=2.45$.

\vspace{.15in}

\newpage

\begin{center}
\begin{tabular}{||c|c||} \hline
$\beta$ & $\rm{f}_{m}$ \\ \hline
$2.40$ & $2.75(6)\times 10^{-2}$ \\ \hline
$2.45$ & $1.95(1)\times 10^{-2}$ \\ \hline
$2.50$ & $1.36(1)\times 10^{-2}$ \\ \hline
\end{tabular}
\end{center}
\begin{center}
{\bf Table I}\\
\end{center}
\vspace{1.0in}

\begin{center}
\begin{tabular}{||c|c|c||} \hline
$\beta$ & $\sigma$ & $\alpha$\\ \hline
2.40 & 0.068(2) & 0.01(1) \\ \hline
2.45 & 0.051(1) & 0.02(1) \\ \hline
2.50 & 0.034(1) & 0.01(1) \\ \hline
\end{tabular}
\end{center}
\begin{center}
\bf{Table II}
\end{center}

\vspace{1.0in}

\vspace{1.0in}

\begin{center}
\begin{tabular}{||c|c|c||} \hline
$\beta$ & $\sigma$ & $\alpha$\\ \hline
2.40 & 0.072(3) & -0.28(2) \\ \hline
2.45 & 0.049(1) & -0.29(1) \\ \hline
2.50 & 0.033(2) & -0.29(1) \\ \hline
\end{tabular}
\end{center}
\begin{center}
\bf{Table III}
\end{center}

\vspace{1.0in}

\begin{center}
\begin{tabular}{||c|c|c|c|c||} \hline
	&\multicolumn{4}{c||}{ max loop size} \\ \hline
$\beta$ & 10 & 20 & 50 & 100 \\ \hline
2.40 & 0.33 & 0.41 & 0.46 & 0.48 \\ \hline
2.45 & 0.38 & 0.46 & 0.51 & 0.54 \\ \hline
2.50 & 0.43 & 0.51 & 0.56 & 0.59 \\ \hline
\end{tabular}
\end{center}
\begin{center}
\bf{Table IV}
\end{center}
\vspace{1.0in}

\begin{center}
\begin{tabular}{||c|c|c|c||} \hline
$\beta$ &$\sigma(\ge 50)$ &$\sigma(\ge 100)$  &$\sigma(\ge 200)$\\ \hline
2.40 & 0.068(3) & 0.066(3)  & 0.063(3)\\ \hline
2.45 & 0.050(3) & 0.049(3)  & 0.044(3)\\ \hline
2.50 & 0.035(2) & 0.034(2)  & 0.029(2)\\ \hline
\end{tabular}
\end{center}
\begin{center}
\bf{Table V}
\end{center}

\newpage

\includegraphics{mono.ps}

\vspace*{8.0in}

\begin{center}
{\bf Figure 1}
\end{center}

\newpage

\includegraphics{comp.ps}

\vspace*{8.1in}

\begin{center}
{\bf Figure 2}
\end{center}

\end{document}